\documentclass[prd,superscriptaddress,twocolumn,nopreprintnumbers,floatfix,nofootinbib]{revtex4}
\usepackage{url,amssymb,amsmath,longtable,rotating,units,wasysym,subfigure,epsfig,multirow,amsbsy,soul,mathrsfs,mathtools}
\usepackage[plainpages=false, colorlinks=true, anchorcolor=blue, linkcolor=blue, citecolor=blue, bookmarks=false]{hyperref}
\usepackage{float}
\usepackage{resizegather}
\usepackage{appendix}
\usepackage{bbold}
\usepackage{graphicx}
\usepackage{mathtools}
\usepackage[dvipsnames]{xcolor}
\usepackage{lipsum}
\usepackage{comment}
\usepackage{breqn}
\usepackage{autobreak}
\usepackage{empheq}

\usepackage{xcolor}
\usepackage{graphicx}
\usepackage{dcolumn}
\usepackage{bm}
\usepackage{color}
\usepackage{ulem,units}
\usepackage{verbatim}

\newcommand{\SPA}{School of Physics and Astronomy, Monash University, Vic 3800, Australia}
\newcommand{\OzGravMonash}{OzGrav: The ARC Centre of Excellence for Gravitational Wave Discovery, Clayton VIC 3800, Australia}

\begin{document}
\title{Did Goryachev et al.\ detect megahertz gravitational waves?}
\author{Paul D. Lasky}
\email{paul.lasky@monash.edu}
\author{Eric Thrane}
\email{eric.thrane@monash.edu}
\affiliation{\SPA}
\affiliation{\OzGravMonash}

\begin{abstract}
\citet{goryachev21} recently announced the detection of ``two strongly significant events'' in their Bulk Acoustic Wave High Frequency Gravitational Wave Antenna. They claim many possibilities for the cause of these events, including high-frequency megahertz gravitational waves. We demonstrate these events are not due to gravitational waves for two reasons. 1) The inferred stochastic gravitational-wave background from these events implies the gravitational-wave energy density of the Universe is $\Omega_{\rm gw}\approx 10^8$, approximately $10^8$ times the closure density of the Universe. 2) The low-frequency gravitational-wave memory signal that accompanies any high-frequency gravitational-wave source visible by the current generation of high-frequency detectors would have been visible by LIGO/Virgo as a transient burst with signal-to-noise ratio $\gtrsim10^6$. The non-detection of such loud memory bursts throughout the operation of LIGO/Virgo rules out the gravitational-wave explanation for the high-frequency events detected by Goryachev et al. We discuss broader implications of this work for the ongoing experimental search for ultra high-frequency (MHz-GHz) gravitational waves.
\end{abstract}

\maketitle

\section{Introduction}
The LIGO/Virgo Scientific Collaborations~\cite{LIGO,Virgo} have announced the detection of more than 50 gravitational-wave observations from merging black holes and neutron stars~\cite{gwtc1,gwtc2,nsbh}. These observatories are sensitive to gravitational waves in the audio band ($\sim\unit[10-1000]{Hz}$). There are numerous experiments across the globe striving to detect gravitational waves in other parts of the spectrum, including a range of instruments looking in the megahertz band~\cite[e.g.,][]{Arvanitaki,goryachev14,holometer}. One of these experiments, the Bulk Acoustic Wave High-Frequency Gravitational Wave Antenna~\cite{goryachev14}, recently provided the data underpinning the detection of two transient events~\cite{goryachev21}. 

\citet{goryachev21} explored a range of possibilities for the origin of these transients---from the banal (e.g., internal processes within their apparatus, cosmic rays, meteor activity) to the extraordinary: domain walls, dark-matter candidates, and gravitational waves. Further exploring the gravitational-wave hypothesis, the authors posit that the events are consistent with the merger of primordial black holes of mass $<\unit[4\times10^{-4}]{M_\odot}$ (about the mass of Saturn). 

In this paper, we show these events were not caused by gravitational waves. We provide two arguments. First, we show that the implied gravitational-wave background energy density is $\approx10^8$ times that of the closure density of the Universe, incompatible with modern cosmology. Second, we show that the lower-frequency gravitational-wave memory signal that necessarily accompanies high-frequency signals~\cite{mcneill17} would have been visible in LIGO/Virgo with signal-to-noise ratio $\gtrsim10^6$. The non-detection of such bursts throughout the lifetime of LIGO/Virgo rules out the hypothesis that Goryachev et al.'s events were due to gravitational waves. We discuss the broader implications of this work on the search for ultra high-frequency (MHz-GHz) gravitational waves at the end of the manuscript.

\section{Stochastic Background}
The dimensionless gravitational-wave energy density in a narrow frequency interval $(f, f + \delta f)$ is~\cite{allen99,Moore15}
\begin{align}
    \Omega_{\rm gw}(f)=\frac{1}{\rho_c}\frac{d\rho_{\rm gw}}{d\ln f}=\frac{2\pi^2}{3 H_0^2}\xi f^2 h_c(f)^2 ,\label{eq:omegaGW}
\end{align}
where $\rho_{\rm gw}$ and $\rho_c$ are the gravitational-wave energy density and the critical energy density required to close the Universe, respectively, $H_0$ is the Hubble parameter, and $\xi$ is the duty cycle of those sources (the fraction of time that a signal is present). The second of the~\citet{goryachev21} sources is stated to have a characteristic strain of $h_c\equiv\sqrt{S_h(f)f}\approx2.5\times10^{-16}$ \cite{locus} where $S_h(f)$ is the strain power spectral density, a $\tau\approx\unit[1]{ms}$ signal duration, and a gravitational-wave frequency of $f_0\approx\unit[5.5]{MHz}$. Given the total observing time was $\approx\unit[5300]{hr}$, the duty cycle is $\xi\approx1\times10^{-10}$. Evaluating Eq.~\ref{eq:omegaGW} gives $\Omega_{\rm gw}(f_0)\approx3\times10^{8}$.

Assuming $\Omega_{\rm gw}(f)$ is approximately constant over the interval $\unit[5-8]{MHz}$ where the two bursts were detected, the integrated energy density from these high-frequency gravitational waves is
\begin{align}
    \Omega_{\text{gw}} = \int d(\ln f) \, \Omega_{\text{gw}}(f) \approx 10^8.
\end{align}
This result can vary by factors of $\approx2$ with different assumptions---and possibly more than that depending on how the characteristic strain depends on the assumed signal duration.
Poisson counting statistics contribute an additional uncertainty factor of $6$ (99\% credibility).
However, even with the most conservative of these factors, $\Omega_{\text{gw}}$ is clearly unphysically large since plausible sources must have $\Omega_{\rm gw}\ll1$ in order to avoid tension with basic cosmology.

\section{Gravitational-wave memory}
The non-isotropic emission of gravitational waves causes a non-linear, hereditary effect known as gravitational-wave memory~\cite{braginsky87,christodoulou91}. Any burst of high-frequency gravitational waves---including for example compact binary mergers~\cite[e.g.,][]{braginsky87}, supernovae~\cite[e.g.,][]{burrows96}, and cusps and kinks on cosmic strings~\cite{jenkins21}---is necessarily accompanied by a lower frequency signal where the strain amplitude spectral density $\sqrt{S_h(f)}\propto1/f$~\cite{mcneill17}. If the high-frequency signals detected by~\citet{goryachev21} were due to gravitational waves, then they would have been accompanied by low-frequency ``orphan memory'' bursts in the LIGO/Virgo observatories~\cite{LIGO,Virgo}.

Converting the quoted characteristic strain $h_c$ into a strain amplitude spectral density implies $\sqrt{S_h(f_0)}=\unit[10^{-19}]{Hz^{-1/2}}$, consistent with Fig.~3 of~\citet{goryachev21}. In~\citet{mcneill17}, we showed that if the bulk acoustic wave experiment could reach their design goal of $\sqrt{S_h(f_0)}=\unit[10^{-22}]{Hz^{-1/2}}$~\cite{goryachev14}, a detected gravitational-wave signal would have a corresponding orphan memory burst in LIGO/Virgo with signal-to-noise ratio $\rho\gtrsim10^3$. Given $S_h\propto\rho$, we estimate the orphan memory signal from the Goryachev et al.\ events to be $\rho\gtrsim10^6$.
Such loud signals would have been detectable in \textit{all} LIGO/Virgo observing runs dating back to the first science run S1 in 2002.
The fact that audio-band gravitational-wave detectors do not regularly detect loud memory bursts is a second sign that the events described by Goryachev et al.\ are not gravitational waves.

\section{Summary}
While \citet{goryachev21} do not claim that the events in their acoustic wave cavity are definitely due to gravitational waves, they propose high-frequency gravitational waves as a serious explanation.
The instrument, which is referred to as a ``high frequency gravitational wave antenna,'' is presented as capable of detecting gravitational waves.
However, the candidate events cannot be due to gravitational waves without posing major challenges to our understanding of general relativity and cosmology.

There is a global effort to detect gravitational waves in the MHz-GHz range---see~\citet{aggarwal20} for a recent review. A number of cosmological and astrophysical sources have been proposed for the emission of gravitational waves in this frequency range~\cite[see][and references therein]{cruise12, aggarwal20}, although it is fair to say these are all speculative. Regardless, our results suggest current detectors are a considerable way from the required sensitivity to probe strain amplitudes not already ruled out from cosmology and orphan memory. For example, given $\Omega_{\rm gw}\propto h_c^2$, the characteristic strain sensitivity of the Bulk Acoustic Wave Gravitational Wave Antenna must improve by approximately four orders of magnitude before a positive detection implies $\Omega_{\text{gw}}\lesssim1$. Table 1 of ~\citet{aggarwal20} provides characteristic strain sensitivities off all current and proposed instruments; the Holometer~\cite{holometer} is the only MHz-GHz detector built with sensitivity such that $\Omega_{\text{gw}}\ll1$ as required by our understanding of the standard model of cosmology.

Following the original submission of our manuscript, Dom\`enech submitted a complementary manuscript to the arXiv~\cite{domenech21}. This work agrees with our general conclusion that the~\citet{goryachev21} events \textit{cannot} be cosmological in nature because the implied energy densities are inconsistent with our understanding of cosmology. They consider the possibility that the events could be due to gravitational waves from `the merging of two planet mass primordial black holes ($\approx\unit[4\times10^{-4}]{M_\odot}$) inside the Oort cloud\ldots,' although noted that the probability of such an event occurring within this volume per year is about $10^{-24}$. While this calculation is enough to cast significant doubt on the primordial black hole scenario, we further note that this scenario necessarily would yield a lower-frequency orphan memory signals unambiguously observable in the LIGO/Virgo observatories.

\section*{ACKNOWLEDGMENTS}
This work is supported through Australian Research Council (ARC) Future Fellowship FT160100112, Centre of Excellence CE170100004, Linkage Grant LE210100002, and Discovery Project DP180103155. 
We are grateful to Ilya Mandel for helpful suggestions on a draft of this manuscript.

\bibliography{GoryachevComment}
\end{document}